\documentclass{PoS}

\title{Baryonic states in supersymmetric Yang-Mills theory}

\ShortTitle{Baryonic states in supersymmetric Yang-Mills theory}

\author{\speaker{Sajid Ali}\\
	Institut f\"ur Theoretische Physik, Universit\"at M\"unster, Wilhelm-Klemm-Str.~9, D-48149 M\"unster\\
	Department of Physics, Government College University Lahore, Lahore 54000, Pakistan\\
	E-mail: \email{sajid.ali@uni-muenster.de}}

\author{Georg Bergner, Camilo Lopez\\
	Theoretisch-Physikalisches Institut, Universit\"at Jena, Max-Wien-Platz 1, D-07743 Jena\\
        E-mail: \email{georg.bergner@uni-jena.de, camilo.lopez@uni-jena.de}}

\author{Henning Gerber, Gernot M\"unster\\
	Institut f\"ur Theoretische Physik, Universit\"at M\"unster, Wilhelm-Klemm-Str.~9, D-48149 M\"unster\\
        E-mail: \email{h.gerber@uni-muenster.de, munsteg@uni-muenster.de}}

\author{Istvan Montvay\\
  	Deutsches Elektronen-Synchrotron DESY, Notkestr.~85, D-22603 Hamburg, Germany
  	E-mail: \email{montvay@mail.desy.de}}

\author{Stefano Piemonte\\
Institute for Theoretical Physics, Universit\"at Regensburg, Universit\"atsstr.~31, D-93053 Regensburg\\
        E-mail: \email{stefano.piemonte@ur.de}}

\author{Philipp Scior\\
Fakult\"at f\"ur Physik Universit\"at Bielefeld Universit\"atsstr.~25, D-33615 Bielefeld\\
        E-mail: \email{scior@physik.uni-bielefeld.de}}

\usepackage{amsmath,amsfonts,amssymb}
\newcommand{\I}{\ensuremath{\mathrm{i}}}
\newcommand{\tr}{\ensuremath{\mathrm{tr}}}
\newcommand{\tu}{\bigtriangleup}
\newcommand{\su}[1]{\ensuremath{\text{SU}(#1)}}
\newcommand{\rpm}{\sbox0{$1$}\sbox2{$\scriptstyle\pm$}
  \raise\dimexpr(\ht0-\ht2)/2\relax\box2 }

\newcommand{\arxiv}[2]{[arXiv:\,\href{http://arxiv.org/abs/#1}{\texttt{#1}} [\texttt{#2}]]}
\newcommand{\arxivold}[1]{[arXiv:\,\href{http://arxiv.org/abs/#1}{\texttt{#1}}\,]}

\abstract{In $\mathcal{N}$=1 supersymmetric Yang-Mills theory the
superpartner of the gluon is the gluino, which is a spin 1/2 Majorana
particle in the adjoint representation of the gauge group. Combining three
gluinos, it is possible to form colour neutral bound states, analogous to
baryons in QCD. The correlation functions of the corresponding baryonic
operators contain a contribution represented by a ``sunset diagram'', and in
addition, unlike in QCD, another contribution represented by a ``spectacle
diagram''. We present first results from an implementation and calculation
of these objects, obtained from numerical simulations of supersymmetric
Yang-Mills theory.}

\FullConference{The 36th Annual International Symposium on Lattice Field Theory - LATTICE2018\\
		22-28 July, 2018\\
		Michigan State University, East Lansing, Michigan, USA.}

\begin{document}


\section{Introduction}

Despite being an extremely successful effective theory, the Standard Model
of particle physics needs some extensions. $\mathcal{N}=1$ supersymmetric
Yang-Mills theory (SYM) is the supersymmetric extension of the gluonic
sector of the Standard Model. The superpartner of the gluon is the gluino.
Gauge invariance together with supersymmetry requires gluinos to transform
according to the adjoint representation of the gauge group \su{N}. SYM is
similar to QCD in some aspects. It is asymptotically free at high energies,
and due to confinement, the low-energy degrees of freedom are expected to be
colour-neutral bound states of gluons and gluinos~\cite{Amati:1988ft}. Their
masses can be computed non-perturbatively using powerful computing machines.

Supersymmetry (SUSY) is the essential feature distinguishing SYM from QCD.
If SUSY is unbroken in the continuum, as non-perturbative studies of the
supersymmetric Ward identities indicate~\cite{Ali:2018fbq}, the bound states
are arranged in supermultiplets, see~\cite{Bergner:2015adz,Ali:2018dnd} for
numerical investigations for gauge groups \su{2} and \su{3}. Effective
actions for the low-energy degrees of freedom in SYM have been constructed
in~\cite{Veneziano:1982ah} and extended
in~\cite{Farrar:1997fn,Farrar:1998rm}. They predict supermultiplets
consisting of mesons, glueballs, and gluino-glue bound states. It is,
however, also possible to construct colour singlet bound states of three
gluinos, analogous to baryons in QCD. These baryonic states are not
described by the low-energy effective
actions~\cite{Veneziano:1982ah,Farrar:1997fn} and it is unknown how they fit
into the supermultiplets. In this paper we investigate these interesting
objects.


\section{$\mathcal{N}=1$ supersymmetric Yang-Mills theory}\label{intro}

The continuum Lagrangian of SYM in Euclidean space-time is given by
\begin{equation}
\mathcal{L} = \frac{1}{4} F^a_{\mu\nu} F^a_{\mu\nu} 
+ \frac{1}{2} \bar{\lambda}^a \gamma_{\mu} (\mathcal{D}_{\mu} \lambda)^a 
+ \frac{m_g}{2} \bar{\lambda}^a \lambda^a,
\end{equation}
where $F^a_{\mu\nu}$ is the non-Abelian field strength tensor built from the
gauge field $A_{\mu}(x)$, which represents the gluon. The gluino field
$\lambda(x)$, the fermion superpartner of the gluon, is minimally coupled to
the gauge field. The gluino satisfies the Majorana condition
\begin{equation}
\overline{\lambda}(x) = \lambda^T(x) \, C\,,
\end{equation}
where $C$ is the charge conjugation matrix. The gluino transforms under the
adjoint representation of the gauge group. The covariant derivative thus
acts as
\begin{equation}
(\mathcal{D}_{\mu} \lambda)^a 
= \partial_{\mu} \lambda^{a} + g_0\,f^a_{bc} A^b_{\mu} \lambda^{c}\,.
\end{equation}

A non-zero mass $m_g$ of the gluino in supersymmetric Yang-Mills theory
causes a soft breaking of supersymmetry. For vanishing gluino mass it is
expected that supersymmetry is unbroken in the
continuum~\cite{Witten:1982df}.

For non-perturbative studies of SYM the theory is regularised on a Euclidean
space-time lattice. The introduction of a space-time lattice regulator
necessarily breaks supersymmetry \cite{Bergner:2009vg}. Models that preserve
part of an extended superalgebra are discussed in
Ref.~\cite{Catterall:2009it}. In our simulations, we use the
``Curci-Veneziano'' lattice action \cite{Curci:1986sm}
\begin{equation}
\label{action}
S = S_g + S_f\, ,
\end{equation}   
where the gauge part ($S_g$) of the full action is given by the standard
Wilson action
\begin{equation}
S_g = \frac{\beta}{2} \sum_x\sum_{\mu\neq\nu}
\left( 1 - \frac{1}{N_c} {\rm Re}\,\tr\, U_{\mu\nu}(x) \right).
\end{equation}
Here $\beta \equiv 2N_c/g_0^2$ is the bare lattice gauge coupling for the
SU($N_c$) gauge field and $U_{\mu\nu}(x)$ is the standard plaquette
variable. The fermion part of the action in equation (\ref{action}) is
\begin{equation}
\begin{array}{rcl}
S_f &\equiv& \frac{1}{2} \sum_x \left\{ \overline{\lambda}^a(x)\lambda^a(x)
- \kappa \sum_{\mu=1}^4 \left[\overline{\lambda}^a(x+\hat{\mu}) 
V_{\mu}^{ab}(x)(1 + \gamma_{\mu}) \lambda^b(x)
+ \overline{\lambda}^a(x) V_{\mu}^{abT}(x) (1 - \gamma_{\mu}) 
\lambda^b(x+\hat{\mu}) \right] \right.\\[2mm]
& & + \left.\frac{\I}{4} g_0 \kappa c_{SW} 
\overline{\lambda}^a(x) \sigma_{\mu\nu} P^{(cl)}_{\mu\nu;ab}(x) \lambda^b(x) 
\right\}.
\end{array}
\end{equation}
Here $\kappa$ is the hopping parameter, which is related to the bare gluino
mass by $\kappa=1/(2m_g+8)$, $\gamma_\mu$ denotes a Dirac matrix, and
$\sigma_{\mu\nu}=-\frac{1}{2}[\gamma_\mu,\gamma_\nu]$. The links
$V_\mu^{ab}(x)$ are the gauge field variables in the adjoint representation,
obtained from links $U_\mu(x)$ in the fundamental representation by
\begin{equation}
V_\mu^{ab}(x) \equiv 
2 \,\tr \left( U^{\dagger}_{\mu}(x) T^a  U_{\mu}(x) T^b \right) ,
\end{equation}
where $T^a$ are the group generators of SU($N_c$). Finally, the last term in
the fermion action contains the clover-symmetrised lattice field strength
tensor:
\begin{equation}
P^{(cl)}_{\mu\nu}(x) = \frac{1}{4a} \sum_{i=1}^{4} \frac{1}{2\I g_0a} 
\bigg( U^{(i)}_{\mu\nu}(x) - U^{(i)\dagger}_{\mu\nu}(x)\bigg)\,.
\end{equation}
The coefficient $c_{SW}$ is tuned up to one-loop order in perturbation
theory to improve the convergence of on-shell observables to the continuum
limit \cite{Musberg:2013foa}.


\section{Baryons in Supersymmetric Yang-Mills theory}

By combining three gluino fields $\lambda(x)$, it is possible to construct a
colourless object, provided all the colour indices are contracted with
indices of invariant colour tensors $t_{abc}$ for $a, b, c = 1, \ldots ,
N_c^2-1$. The corresponding particles are analogous to baryons in Quantum
Chromodynamics (QCD), and will therefore also be called ``baryons''. Due to
the fact that gluinos are in the adjoint representation of the gauge group,
these objects can be constructed for both SU(2) and SU(3).

Interpolating fields for baryons will generally be of the form
\begin{equation}
W(x) = t_{abc} \lambda_a(x) \big( \lambda^T_b(x) \Gamma \lambda_c(x) \big)
\end{equation}
with an invariant colour tensor $t_{abc}$ and a matrix $\Gamma$ in Dirac
space. In the case of \su{2} one has $t_{abc}=\epsilon_{abc}$ and $\Gamma$
has to be symmetric. For \su{3} the $t_{abc}$ can be chosen as the
completely antisymmetric structure constants $f_{abc}$ or as the completely
symmetric tensor $d_{abc}$. The spin matrix $\Gamma$ then has to have the
opposite symmetry type in order that $W \neq 0$. Depending on the choice of
$\Gamma$, the baryon field $W(x)$ carries either spin 1/2 only or both spin
1/2 and spin 3/2. These parts can be singled out with suitable projection
operators~\cite{Leinweber:2004it}. A possibility, which we considered, is
the choice $t_{abc}=\epsilon_{abc}$ and $\Gamma=C\gamma_4$ for gauge group 
$\su{2}$.

The conjugate field is
\begin{equation}
\overline{W}(x) = -\big(C\,W(x)\big)^T .
\end{equation}
The masses of baryons can be determined from the corresponding correlation
functions
\begin{align}
D(x,y) &= \langle W(x) \overline{W}(y) \rangle \nonumber\\
D^{\alpha \delta}(x,y) &= \langle W^{\alpha}(x) \overline{W}^{\delta}(y) 
\rangle \nonumber\\
&= - \langle W^{\alpha}(x) C^{\delta \alpha'} W^{\alpha'}(y) \rangle \nonumber\\
&= - t_{abc} t_{a'b'c'} \Gamma^{\beta \gamma}\Gamma^{\beta' \gamma'} 
C^{\delta \alpha'}\label{bcorr}
\langle
\lambda_a^{\alpha}(x)
\lambda_b^{\beta}(x)
\lambda_c^{\gamma}(x)
\lambda_{a'}^{\alpha'}(y)
\lambda_{b'}^{\beta'}(y)
\lambda_{c'}^{\gamma'}(y)
\rangle .
\end{align}
The fermionic expectation value can be decomposed into gluino propagators by
means of Wick's theorem. In SYM, in contrast to QCD, every pair of fermion
fields contributes due to the Majorana nature of the gluino. In order to
obtain the Wick contractions, it is convenient to express the correlation
function in terms of six $\lambda$, by converting factors $\bar{\lambda}$
into $\lambda$ by means of the Majorana condition. The Wick contractions are
\begin{equation}
\langle \lambda^{\alpha}_a(x) \lambda^{\beta}_b(y) \rangle
= K^{\alpha \beta}_{ab}(x,y) = -(\tu(x,y) C)^{\alpha \beta}_{ab},
\end{equation}
where $\tu = Q^{-1}$ is the fermion propagator. Taking into account the
fermionic signs, we get 15 terms:
\begin{align}
D^{\alpha\alpha'}(x,y) = &\,t_{a'b'c'} t_{abc} \Gamma^{\beta\gamma} 
\Gamma^{\beta'\gamma'} \times \{\nonumber\\
& + 2 \tu^{\alpha \alpha'}_{a a'}(x,y) \tu^{\beta \beta'}_{b b'}(x,y)
\tu^{\gamma \gamma'}_{c c'}(x,y)\nonumber\\
& + 4 \tu^{\alpha \beta'}_{a b'}(x,y) \tu^{\beta \gamma'}_{b c'}(x,y) 
\tu^{\gamma \alpha'}_{ca'}(x,y)\nonumber\\
& + 2 \tu^{\alpha \beta}_{ab}(x,x)   \tu^{\delta \alpha'}_{ca'}(x,y)
\tu^{\delta' \beta'}_{c' b'}(y,y) C^{\gamma \delta} 
C^{\gamma' \delta'}\nonumber\\
& + 4 \tu^{\alpha \beta}_{ab}(x,x)   \tu^{\beta' \gamma}_{b'c}(y,x) 
\tu^{\gamma' \alpha'}_{c'a'}(y,y)\nonumber\\
& + 1 \tu^{\alpha \alpha'}_{aa'}(x,y)\tu^{\beta \delta}_{bc}(x,x) 
\tu^{\delta' \beta'}_{c'b'}(y,y) C^{\gamma\delta} 
C^{\delta'\gamma'}\nonumber\\
& + 2 \tu^{\alpha \delta'}_{ac}(x,y) \tu^{\beta \delta}_{bc}(x,x) 
\tu^{\beta' \alpha'}_{b'a'}(y,y) C^{\gamma\delta} C^{\gamma'\delta'}
\}.
\end{align}
\begin{figure}[b]
\centering
\includegraphics[width=0.3\textwidth,clip]{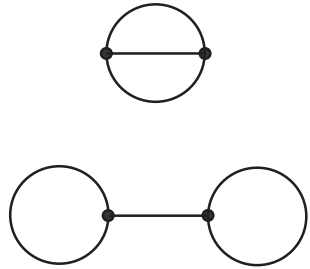}
\caption{Contributions to the propagator of a three-gluino state. The second
contribution, which we call ``spectacle graph'' comes from contractions of
the form $\lambda(x)\lambda(x)$, which are allowed for Majorana fermions.}
\label{Sset_Spect}
\end{figure}
Depending on the topology of the associated fermion line diagrams, the
contributions can be divided into two types called ``Sunset'' and
``Spectacle'', which are represented in Fig.\ref{Sset_Spect}. 

The corresponding algebraic expressions are given by
\begin{align}\label{Sun}
D_{\text{sunset}}(x,y) = &\,t_{a'b'c'} t_{abc} \Gamma^{\beta\gamma} 
\Gamma^{\beta'\gamma'} P_{\rpm}^{\alpha\alpha'} \times \{\\
& + 2 \tu^{\alpha \alpha'}_{a a'}(x,y) \tu^{\beta \beta'}_{b b'}(x,y)
\tu^{\gamma \gamma'}_{c c'}(x,y)\\
& + 4 \tu^{\alpha \beta'}_{a b'}(x,y) \tu^{\beta \gamma'}_{b c'}(x,y) 
\tu^{\gamma \alpha'}_{ca'}(x,y)\}
\end{align}
\begin{align}\label{Spec}
D_{\text{spectacle}}(x,y) = &\,t_{a'b'c'} t_{abc} \Gamma^{\beta\gamma} 
\Gamma^{\beta'\gamma'} P_{\rpm}^{\alpha\alpha'} \times \{\nonumber\\
& + 2 \tu^{\alpha \beta}_{ab}(x,x)   \tu^{\delta \alpha'}_{ca'}(x,y)
\tu^{\delta' \beta'}_{c' b'}(y,y) C^{\gamma \delta} C^{\gamma' \delta'}
\nonumber\\
& + 4 \tu^{\alpha \beta}_{ab}(x,x)   \tu^{\beta' \gamma}_{b'c}(y,x) 
\tu^{\gamma' \alpha'}_{c'a'}(y,y)\nonumber\\
& + 1 \tu^{\alpha \alpha'}_{aa'}(x,y)\tu^{\beta \delta}_{bc}(x,x) 
\tu^{\delta' \beta'}_{c'b'}(y,y) C^{\gamma\delta} C^{\delta'\gamma'}
\nonumber\\
& + 2 \tu^{\alpha \delta'}_{ac}(x,y) \tu^{\beta \delta}_{bc}(x,x) 
\tu^{\beta' \alpha'}_{b'a'}(y,y) C^{\gamma\delta} C^{\gamma'\delta'}\}
\end{align}
In contrast to mesonic or glueball correlation functions, the baryonic
correlation functions are not symmetric or antisymmetric with respect to
Euclidean time, if no parity projection is included, see e.\,g.~\cite{MM}.
Therefore we include the projection onto parity eigenstates by means of the
parity projector $P_{\rpm}$, which is defined as $P_{\rpm} = \frac 12 (1
\rpm \gamma_4)$ for zero momentum.

We have implemented the contributions to the baryonic correlation function
into the measurement code of our SYM collaboration. The numerically quite
expensive part is the evaluation of the spectacle piece. We have calculated
it by means of combining the exact contribution of the lowest eigenmodes of
the Wilson-Dirac matrix with the stochastic estimator technique.

\begin{figure}[b]
\centering
\includegraphics[width=0.8\textwidth,clip]{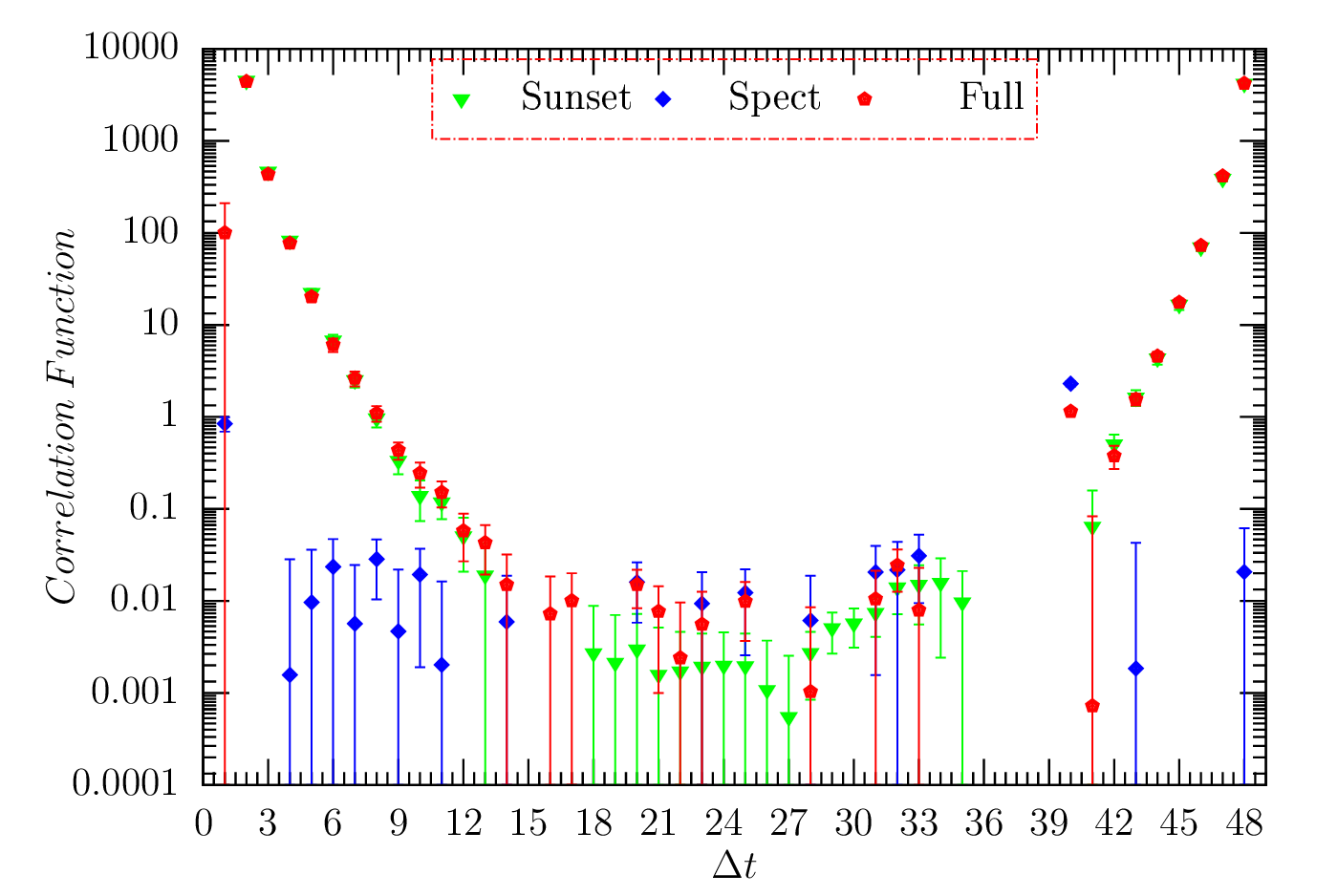}
\caption{Numerical results for correlation functions at each time slice for
gauge group SU(2), lattice volume, $V = 24^3 \cdot 48$, $\beta = 1.75$ and
$\kappa = 0.14925$.}
\label{corr}
\end{figure}
\begin{figure}[t]
\centering
\begin{minipage}{0.45\textwidth}
    \centering
    \includegraphics[width=1.1\linewidth,clip]{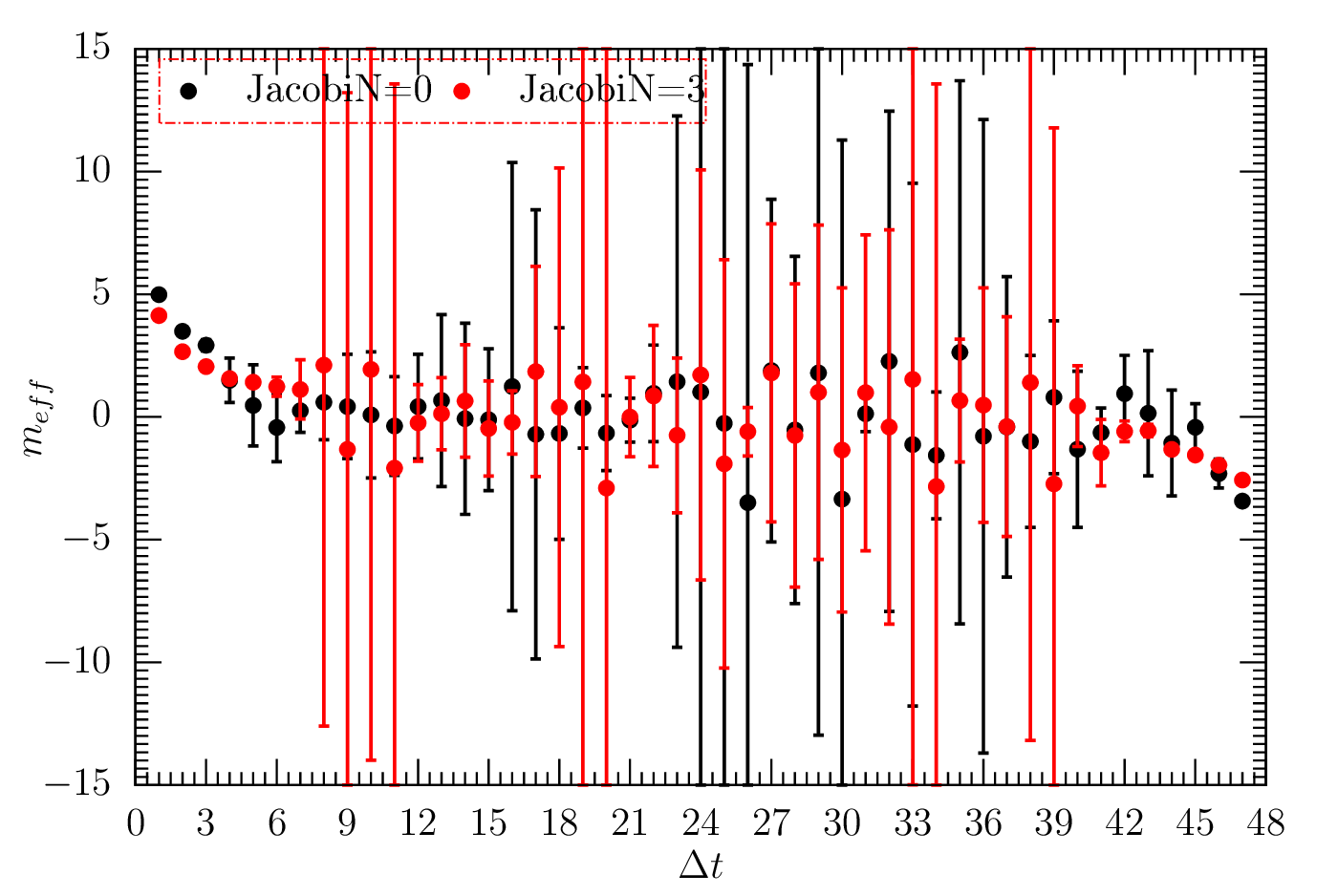}
    \includegraphics[width=1.1\linewidth,clip]{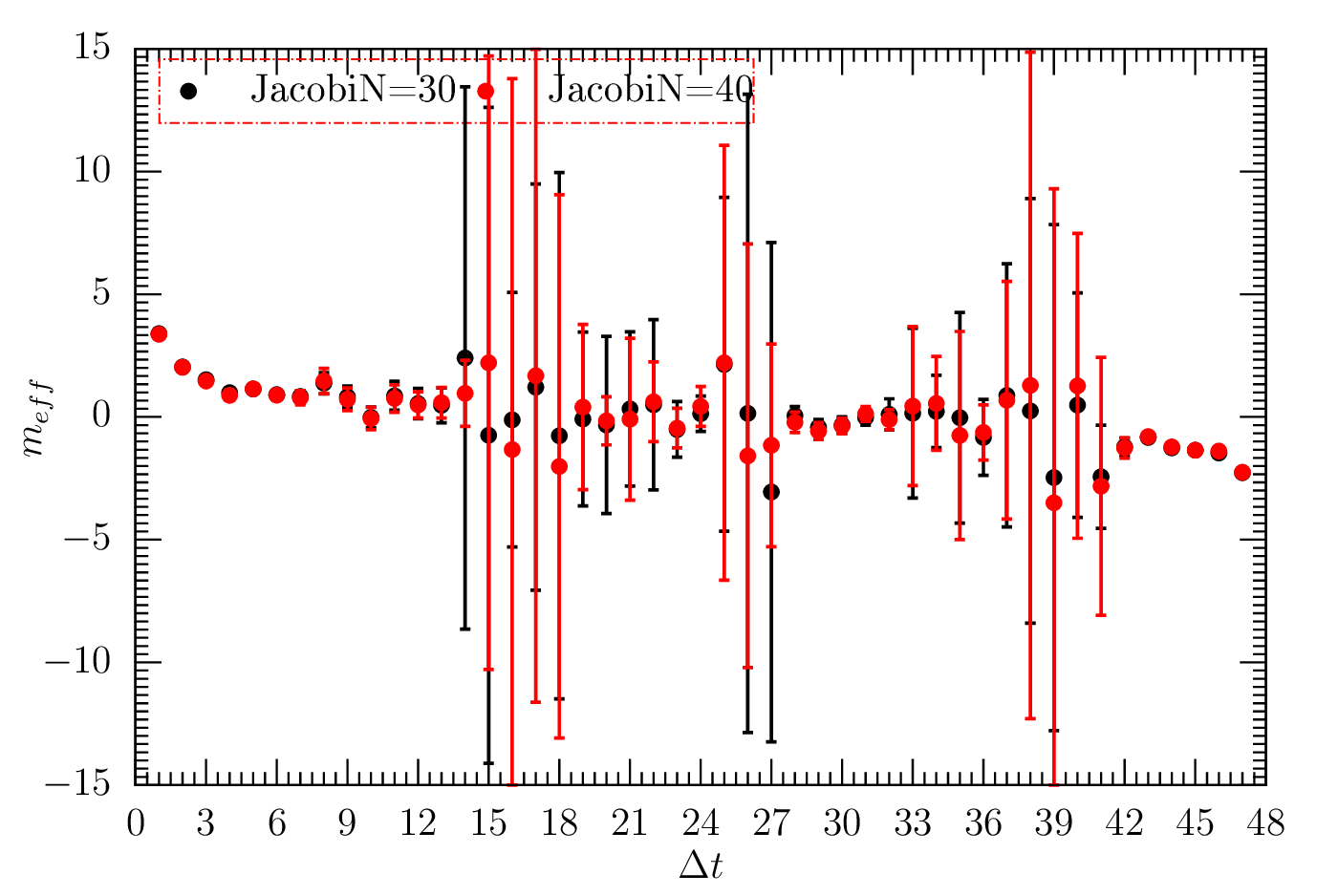}
\end{minipage}\hspace{5mm}
\begin{minipage}{0.45\textwidth}
    \centering
    \includegraphics[width=1.1\linewidth,clip]{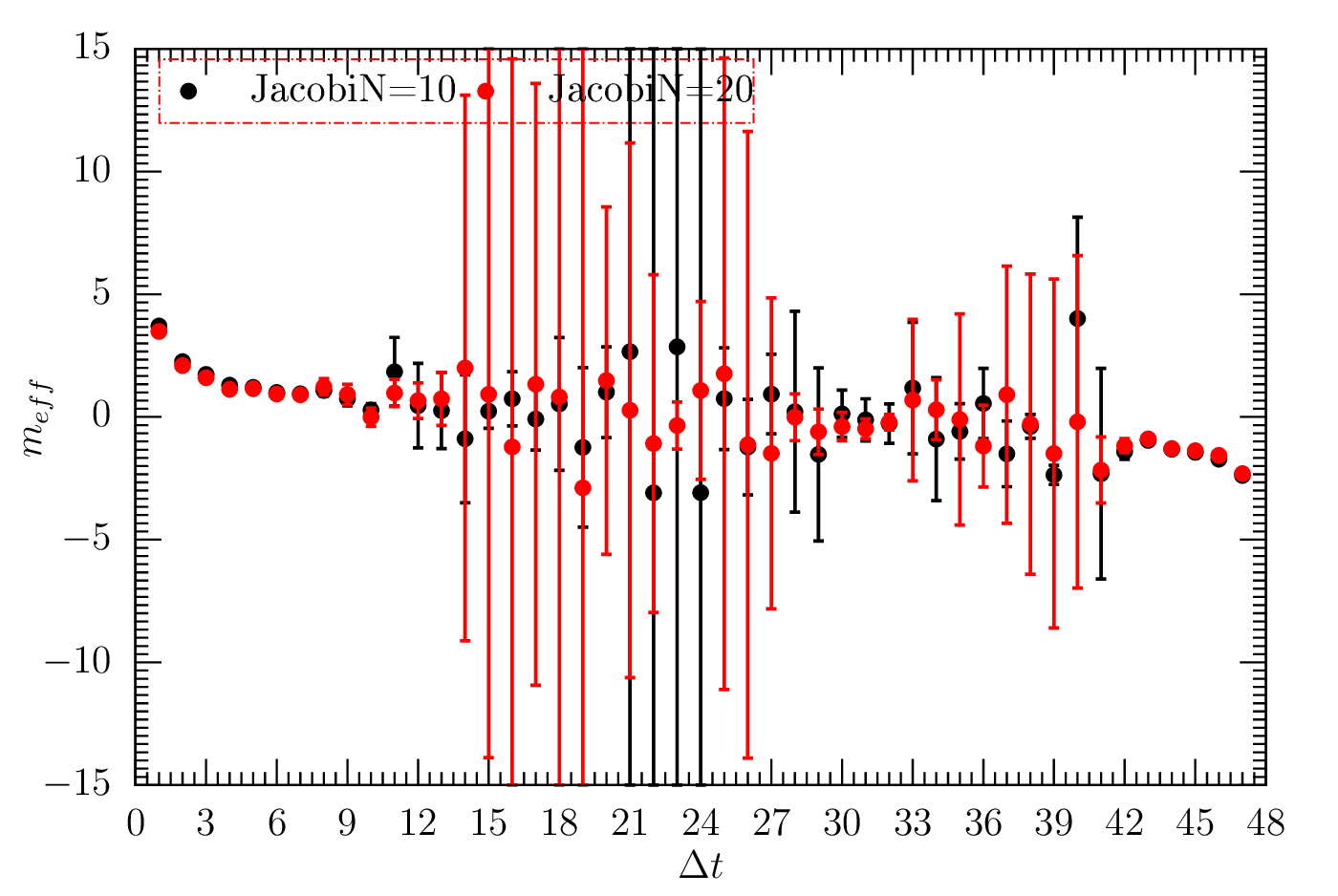}
    \includegraphics[width=1.1\linewidth,clip]{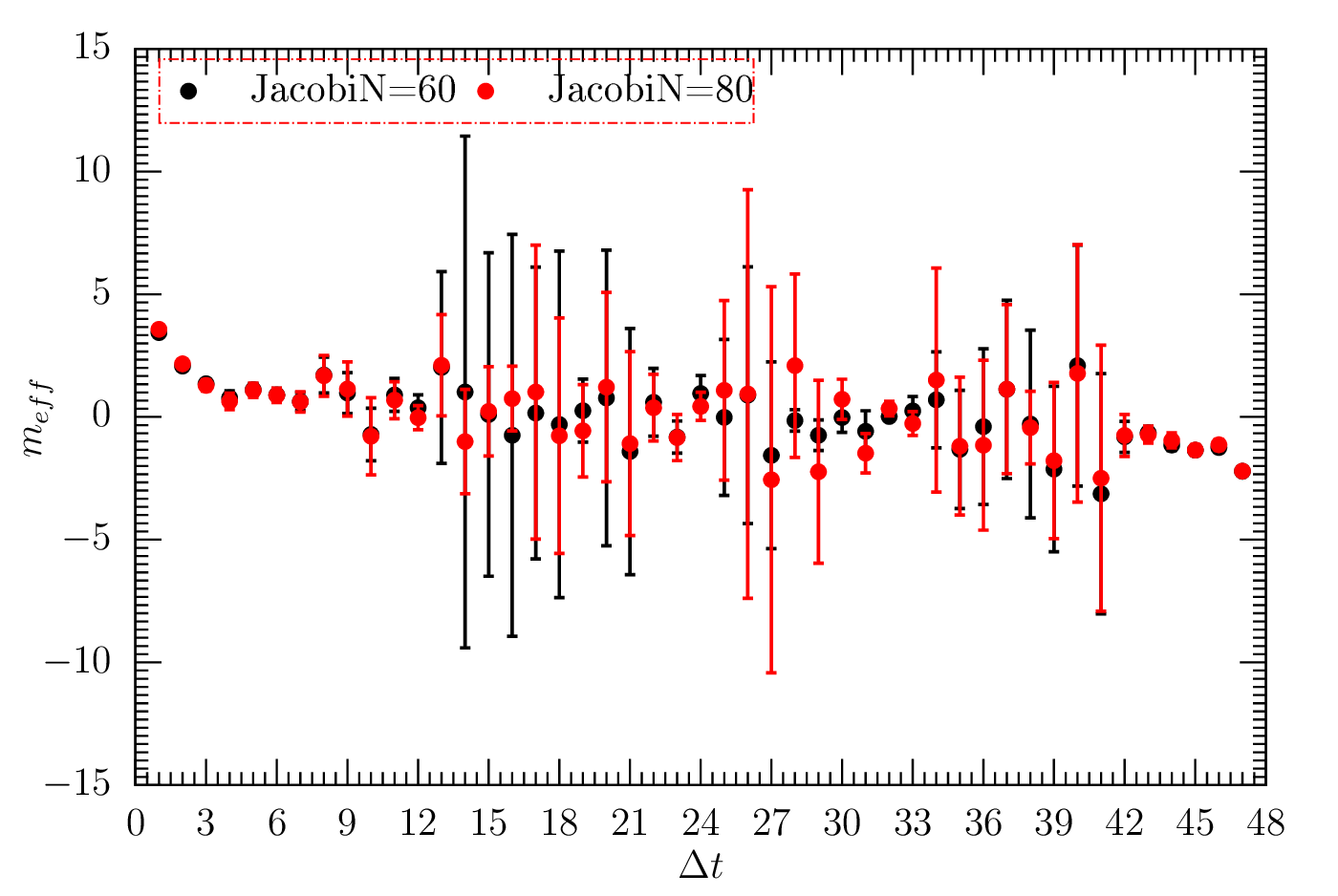}
\end{minipage}
\caption{Effective mass plots for fixed APE pre-smearing with level 40 and
different levels of Jacobi smearing of the Dirac fermion
matrix.}
\label{m_eff}
\end{figure}

Presently some preliminary results have been obtained. Fig.~\ref{corr} shows
the two pieces of the baryonic correlation function and its sum for gauge
group \su{2}. In this case the sunset piece appears to dominate the
correlation function for not too large times. Fig.~\ref{m_eff} shows the
effective mass for different smearings.


\section{Conclusion and outlook}

In this paper we constructed the correlation function for a three-gluino
state in supersymmetric Yang-Mills theory and presented preliminary
estimates of its mass using Monte Carlo simulation at non-zero gluino mass
and at finite inverse gauge coupling $\beta$. With better numerical results
we expect to be able to perform chiral and continuum extrapolations in the
future, and to find out how the baryonic states fit into the supermultiplet
found in Ref.~\cite{Bergner:2015adz}.


\section*{Acknowledgements}

The authors gratefully acknowledge the Gauss Centre for Supercomputing
e.~V.\ (www.gauss-centre.eu) for funding this project by providing computing
time on the GCS Supercomputer JUQUEEN and JURECA at J\"ulich Supercomputing
Centre (JSC) and SuperMUC at Leibniz Supercomputing Centre (LRZ). Further
computing time has been provided on the compute cluster PALMA of the
University of M\"unster. This work is supported by the Deutsche
Forschungsgemeinschaft (DFG) through the Research Training Group ``GRK 2149:
Strong and Weak Interactions-from Hadrons to Dark Matter''. G.~Bergner\
and C. Lopez acknowledge support from the Deutsche Forschungsgemeinschaft (DFG) under
Grant No.\ BE 5942/2-1. S.~Ali acknowledges financial support from the
Deutsche Akademische Austauschdienst (DAAD).


\end{document}